\documentclass{INTERSPEECH2023}

\interspeechcameraready

\usepackage{hyperref}
\usepackage[hyphenbreaks]{breakurl}
\hypersetup{
   breaklinks=true,   
   pdfusetitle=true,  
}

\usepackage{color}
\definecolor{codegreen}{rgb}{0,0.6,0}
\definecolor{codegray}{rgb}{0.5,0.5,0.5}
\definecolor{codepurple}{rgb}{0.58,0,0.82}
\definecolor{backcolour}{rgb}{1.00,1.00,1.00}

\title{Streaming Parrotron for on-device speech-to-speech conversion}
\name{Oleg Rybakov, Fadi Biadsy, Xia Zhang, Liyang Jiang, Phoenix Meadowlark, Shivani Agrawal}
\address{Google Research}
\email{\{rybakov,biadsy,xiaz,jiangliyang,meadowlark,shivaniagrawal\}@google.com}

\newcommand{\sGL}{{\it GL}}
\newcommand{\nGL}{{\it nGL}}
\newcommand{\sDec}{{\it sDec}}

\newcommand{\sMelGAN}{{\it MG}}

\usepackage{setspace}

\begin{document}

\setstretch{0.92}

\maketitle
 
\begin{abstract}
We present a fully on-device streaming Speech2Speech conversion model that normalizes a given input speech directly to synthesized output speech.  Deploying such a model on mobile devices pose significant challenges in terms of memory footprint and computation requirements.  We present a streaming-based approach to produce an acceptable delay, with minimal loss in speech conversion quality, when compared to a reference state of the art non-streaming approach. Our method consists of first streaming the encoder in real time while the speaker is speaking. Then, as soon as the speaker stops speaking, we run the spectrogram decoder in streaming mode along the side of a streaming vocoder to generate output speech. To achieve an acceptable delay-quality trade-off, we propose a novel hybrid approach for look-ahead in the encoder which combines a look-ahead feature stacker with a look-ahead self-attention. We show that our streaming approach is $\approx$2x faster than real time on the Pixel4 CPU.
\end{abstract}

\section{Introduction}
\label{sec:intro}

With the increase in computational resources of mobile devices and advances in speech modeling in recent years, on-device inference has become an important research topic. Automatic Speech Recognition (ASR), Text to Speech (TTS), and  machine translation models, for example, have been ported to run locally on mobile devices with considerable success~\cite{ASR_STREAM, STREAM_ON_DEVICE, DeviceTTS, ON_DEVICE_TTS_STREAM_DECODE}. In this work, we tackle the difficulties of running end-to-end attention-based Speech-To-Speech (STS) models, which directly map input spectrograms to output spectrograms, on mobile devices.  STS models have several applications including STS translation~\cite{SPEECH_TO_SPEECH_MT}, speech conversion and normalization~\cite{Parrotron,PARROTRON_CONFORMER}. 

Our work focuses on Parrotron~\cite{Parrotron}, a STS conversion model that has shown to successfully convert atypical speech from speakers with speech impairments (due to, for example, deafness, Parkinsons, ALS, Stroke, etc.) {\em directly} to fluent typical speech~\cite{SubmodelFramework}. 
To achieve optimal results, Parrotron is adapted for each dysarthric speaker, producing a Submodel for each speaker.~\cite{SubmodelFramework, PARROTRON_DEMO}
Running such a personalized model locally on a device has substantial practical advantages over server-based models, including: scalability, inference without internet connectivity, reduced latency, and privacy.
 
Unlike TTS and ASR, {\em both} Parrotron's input and output are  acoustic frames -- i.e., significantly longer input-output sequences and the model requires more parameters.  Due to these extra challenges, running inference on-device, the model size has to be sufficiently smaller to fit in RAM, and the model must run with limited compute resources.  We have shown that running all the components of the model in non-streaming mode substantially increases the delay between the time the speaker stops speaking and the time the user obtains the converted speech. Controlling the latency of this model is crucial for natural human-to-human communication, when the system is the only mode of communication for  speakers with speech impairments, for example.  
Parrotron is an encoder-decoder model. There are several approaches for streaming such a model. One is to stream the encoder and decoder simultaneously so that the model generates new speech while the user is speaking \cite{STREAM_ASR_MT_TTS}, or generates translated text while the user is speaking~\cite{SimulST_SPEECH_TO_SPPECH}. Alternatively, for speech conversion, we propose to stream the encoder while the user is speaking, and then stream the decoder to generate new speech afterwards, similar to the way the decoder runs in this TTS~\cite{ON_DEVICE_TTS_STREAM_DECODE}. We suspect that this option is well suited for a face-to-face dialog, especially for dysarthric input speech conversion \cite{PARROTRON_CONFORMER}. The reader is encouraged to view demo~\cite{PARROTRON_DEMO}. We present the following contributions:

\setlist{nolistsep}
\begin{itemize}[noitemsep]
    \item A real time on-device streaming STS conversion model where the encoder runs in real-time while the speaker is speaking, and as soon as the user stops speaking, the decoder starts generating output audio.
    \item A novel approach for streaming hybrid lookahead Conformer encoder which outperforms conventional techniques in terms of delay and accuracy, using a combination of lookahead local self-attention and stacker lookahead.
    \item A comprehensive analysis and comparison of several streaming approaches: streaming decoder, streaming causal and non-causal encoders, streaming vocoders and quantization in terms of latency, delay, accuracy, model size and memory footprint.
\end{itemize}

\section{Model architectures} \label{methods}

\subsection{Baseline Parrotron model}\label{base}
\begin{figure}[t]
    \centering
    \includegraphics[width=0.9\columnwidth]{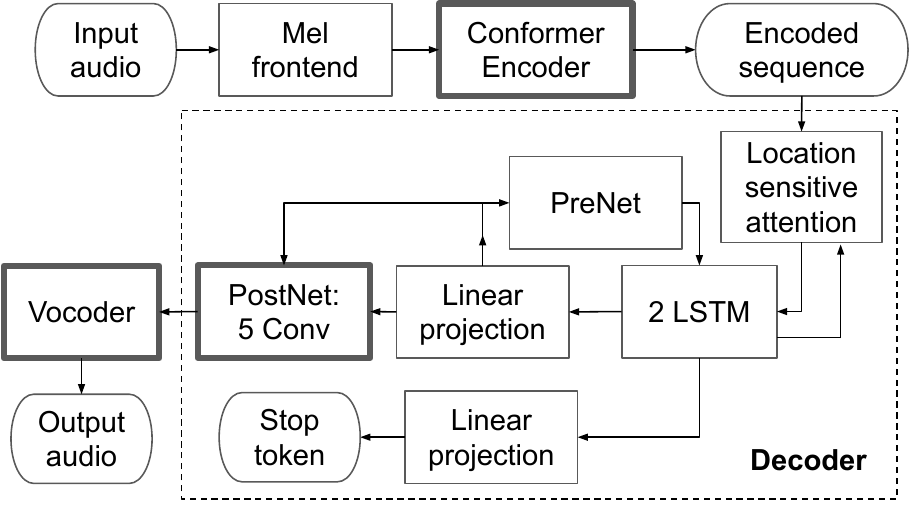}
    \caption{Conformer Parrotron model.}
    \label{fig:parrotron_inference}
\end{figure}

Baseline Parrotron model (we call it \textit{Base} model)~\cite{ PARROTRON_CONFORMER} is composed of a Mel frontend, Conformer encoder,  and an auto-regressive LSTM decoder with cross attention, followed by a vocoder to synthesize a time-domain waveform, as shown in Figure~\ref{fig:parrotron_inference}. \textit{Base} model~\cite{PARROTRON_CONFORMER} takes the whole input speech and processes it end to end. The entire input spectrogram is first encoded and stored in the 'Encoded sequence' block on Figure~\ref{fig:parrotron_inference}. Then, at every decoder step this sequence is processed by a cross 'Location-sensitive attention' block~\cite{TTS_DECODER}. The prediction from previous time steps are processed by a PreNet (two fully connected layers, 256 units each). The PreNet's output is concatenated with the cross-attention context and passed to two unidirectional LSTM layers (1024 units). The LSTM output and the attention context are concatenated and passed to 'Linear projection', which predicts two frames, each of which is 12.5ms, per call. These frames are processed by a PostNet block (5 convolution layers) and added as residual connection to the output of 'Linear projection'. The final decoder's output is two linear spectrogram frames. Finally, the output frames are passed through a 'Vocoder'(in \textit{Base}~\cite{PARROTRON_CONFORMER} model they use non streaming Griffin Lim) block to generate a time-domain waveform of the converted speech.

Conformer encoder in \textit{Base} model is composed of a sequence of layers: 2 conformer blocks; 1 causal Stacker block; 2 conformer blocks; 1 causal Stacker block followed by 13 conformer blocks. In this paper all models have 17 Conformer blocks with total number of parameters of a model 168M as in \textit{Base}~\cite{PARROTRON_CONFORMER}.

\begin{figure}[t]
    \centering
    \includegraphics[width=0.9\columnwidth]{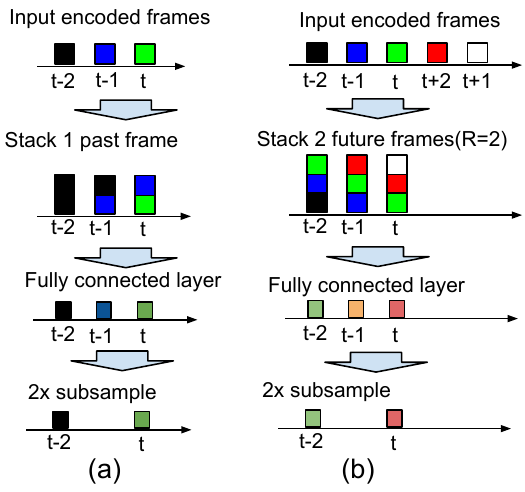}
    \caption{Causal (a) and lookahead (b) Stackers.}
    \label{fig:stacker}
    \vspace{-4mm}
\end{figure}

Causal Stacker block diagram is shown on Figure~\ref{fig:stacker} (a). Input encoded frames (with 512 states), labeled by different colors (on Figure~\ref{fig:stacker} (a)), are processed by several steps: at time \textit{t}, it stacks one past frame from time \textit{t-1}, then pass them through fully connected layer which project them back to original 512 dimensions, then apply 2x subsampling in time dimension to reduce further computations. 

One conformer block~\cite{CONFORMER} is a sequence of fully connected, multi-head full self attention, causal convolution (32x1 kernel size), fully connected and normalization layers. Every layer has a residual connection (except layer normalization). Full self attention layer (with 512 states, 8 attention heads) at time \textit{t} computes self attention over all \textit{T} input frames as shown on Figure~\ref{fig:atention} (c). That is why this model can not run in real time streaming mode: it needs to wait until all frames are available for computation. As a result it has the worst \textit{total delay}. \textit{Total delay} is a difference between the time when a user stops talking and time when output audio (generated by the model) is played back to the user. It is a user-facing metric, so one of the goals of this paper is to find a model with accuracy similar to the \textit{Base} model but with much less \textit{total delay}.

In section~\ref{baseline_experiments} we show that \textit{Base} model has 7 seconds \textit{total delay} of processing 10 seconds of input audio on Pixel4. It makes it unusable for real time user experience. That is why in the next section~\ref{causal} we present the streaming \textit{Causal Base} Parrotron model which addresses this issue. In sections \ref{lsa} and \ref{lsa_ls} we propose further improvements of \textit{Causal Base} model with better \textit{total delay} accuracy trade-off.

\subsection{Streaming \textit{Causal Base} model}\label{causal}

Streaming \textit{Causal Base} model has the similar topology with non streaming \textit{Base} model (shown on Figure~\ref{fig:parrotron_inference}), with several key differences in blocks highlighted by thick line: Conformer Encoder, PostNet, and Vocoder blocks (these blocks are running in streaming mode). The streaming Parrotron model runs encoder in real time streaming mode while the speaker is speaking.  As soon as the speaker stops speaking we have the whole encoded speech available in the block "Encoded sequence" as shown on Figure~\ref{fig:parrotron_inference}. Then we run the spectrogram decoder in streaming mode along the side of a streaming vocoder to generate output speech in real time.

The standard approach of converting a model to streaming mode is to make all modules causal. So in the Decoder block (on Figure~\ref{fig:parrotron_inference}) we use causal convolutions in PostNet and call this decoder streaming decoder {\sDec}. Output of streaming decoder is processed by streaming vocoder (described in \cite{VOCODER_GL_MELGAN}) which generates output audio as shown on Figure~\ref{fig:parrotron_inference}. 

\begin{figure}[t]
    \centering
    \includegraphics[width=0.9\columnwidth]{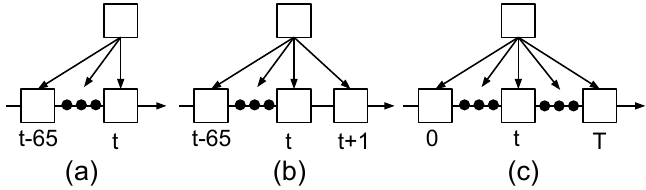}
    \caption{Self attention diagrams: Local causal (a) local lookahead with right context R=1 (b) and full self attention (c).}
    \label{fig:atention}
    \vspace{-4mm}
\end{figure}

We also replace the Conformer encoder (on Figure~\ref{fig:parrotron_inference}) by the Causal conformer encoder. It has the same topology with the Conformer Encoder of the \textit{Base} model (described in section~\ref{base}), but full self attention of the Conformer block is replaced by local causal self attention. Local causal self attention at time t computes local self attention over the previous 65 frames, as shown on Figure \ref{fig:atention}. Causal Conformer encoder has zero delay, so by the time the user stops talking we have all encoded frames stored in "Encoded sequence" block (on Figure~\ref{fig:parrotron_inference}) and decoder can start streaming output audio. We hypothesize that such an approach has significant quality implications. That is why in the next sections we propose a new streaming Conformer encoders which can run in streaming mode at real time and at the same time have minimum quality impact.

\subsection{Streaming lookahead self attention model(\textit{LSA})} \label{lsa}

\begin{figure}[t]
    \centering
    \includegraphics[width=1.0\columnwidth]{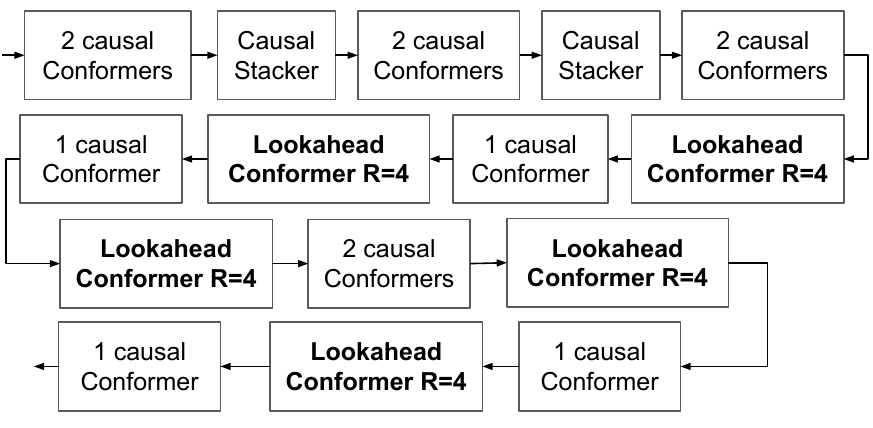}
    \caption{Streaming lookahead Conformer encoder.}
    \label{fig:streaming_lookahead_encoder}
    \vspace{-4mm}
\end{figure}

To improve accuracy of the \textit{Causal Base} model we propose to replace "Conformer Encoder" by streaming "lookahead Conformer encoder" shown on Figure~\ref{fig:streaming_lookahead_encoder}. We call this model \textit{LSA}(lookahead self attention). Lookahead conformer block has the same topology as the standard Conformer block, but full self attention is replaced with lookahead local self attention. It is a popular approach for modeling lookahead, described in \cite{STREAM_TRANSFORMER_LOOK_AHEAD, STREAM_NON_STREAM}, which include future/right hidden-states to improve quality over a fully casual model. For example on Figure~\ref{fig:atention} (b) lookahead local self attention at time t computes self attention over the last 65 frames and future R frames (R = 1). So it is a trade-off between a fully causal (optimal delay) and non-causal encoder (optimal quality). Causal Conformer block is described in section~\ref{causal}. 

\subsection{Streaming lookahead self-attention with a lookahead stacker model(\textit{LSA\_LS})} \label{lsa_ls}
\begin{figure}[t]
    \centering
    \includegraphics[width=0.9\columnwidth]{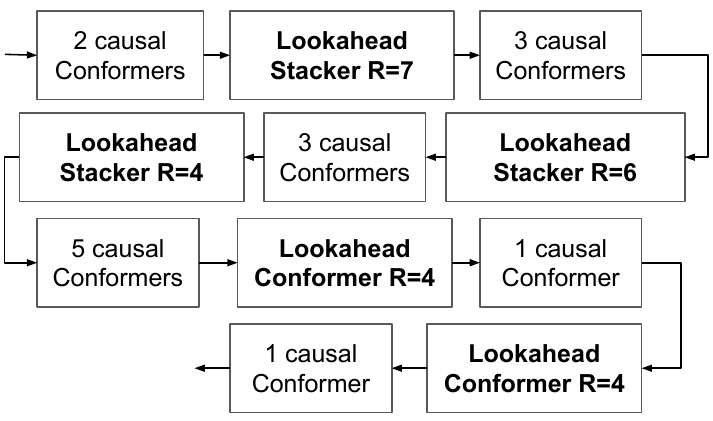}
    \caption{Streaming hybrid lookahead Conformer encoder.}
    \label{fig:streaming_encoder}
\end{figure}

In this section we hypothesize that with the same \textit{total delay} budget it is possible to further improve accuracy of \textit{LSA} model. So in the model described in section~\ref{lsa}  we propose to replace streaming lookahead Conformer encoder (shown on Figure~\ref{fig:streaming_lookahead_encoder}) by streaming hybrid lookahead Conformer encoder shown on Figure~\ref{fig:streaming_encoder}. The key difference with the "lookahead Conformer encoder" is that causal Stacker is replaced by lookahead Stacker. For example, a diagram of lookahead Stacker with R=2 is shown on Figure~\ref{fig:stacker} (b). Input encoded frames (with 512 states) are processed by several steps: at time t, it stacks two future frames from time t+1 and t+2, then pass them through fully connected layer which project them back to original 512 dimensions, then apply 2x subsampling in time dimension to reduce further computations. Since each hidden-state now includes information from the future, we hypothesize that causal conformer blocks is more likely to make use of this information in combination with a lookahead conformer blocks as opposed to having them modeled by the lookahead conformer blocks independently, risking low attention weights(as in \textit{LSA} model). The new streaming hybrid lookahead Conformer encoder is shown on Figure~\ref{fig:streaming_encoder}. It is built using the following stack of layers: 2 causal Conformer blocks; lookahead Stacker block(with right context R=7); 3 causal conformer blocks; lookahead Stacker block(with right context R=6); 3 causal conformer blocks; lookahead Stacker block(with right context R=4); 5 causal Conformer blocks; 1 lookahead Conformer block (with right context R=4); 1 causal Conformer block; 1 lookahead Conformer block (with right context R=4) followed by 1 causal Conformer block. Note that this encoder uses three Stacking layers, thus it reduces the sampling rate by 8$\times$. We call such a model \textit{LSA\_LS} (lookahead self-attention with a lookahead stacker).

\section{Experimental results and discussion} \label{methods}

\subsection{Datasets and key metrics}

All our models are trained on a parallel corpus, where the input utterances are from the entire set of the Librispeech~\cite{Librispeech} (960 hours of training data composed of “train-clean-100”, “train-clean-360”, “train-other-500”) and the corresponding target utterances are the synthesis of their manual transcripts, using Google’s Parallel WaveNet-based single-voice TTS~\cite{ParallelWaveNet}. We use training recipe described in \cite{PARROTRON_CONFORMER}, model training on 16 V3 TPUs takes 4 days. Parrotron effectively builds a many-to-one voice conversion system that normalizes speech from arbitrary speakers to a single canonical voice. 

For WER evaluation, we use the Librispeech test-clean data set as in~\cite{PARROTRON_CONFORMER}. To evaluate the quality of our speech conversion systems, we use Google's state-of-the-art ASR to automatically transcribe the converted speech and then calculate and compare Word Error Rates (WERs) across systems, as in ~\cite{Parrotron, PARROTRON_CONFORMER}. As in~\cite{Parrotron} we use WER as measure of speech intelligibility. Parrotron allows users to communicate not only with humans but rather with any speech-enabled devices, such as Google Home, Amazon Alexa, Siri. Thus ASR WER is a crucial metric to evaluate given the application. 

Another important user facing metric is a \textit{total delay} described in section~\ref{base}. All benchmarks are done with TFLite~\cite{TFL} single-threaded on a Pixel4 CPU, using~\cite{Lingvo} to build our models.

\subsection{Non streaming \textit{Base} model vs streaming models} \label{baseline_experiments}

Due to out-of-memory, we are unable to benchmark \textit{Base} model~\cite{PARROTRON_CONFORMER} end-to-end on Pixel4~\cite{Pixel4}. Therefore, we divide the model into three parts: encoder, decoder and vocoder and benchmark them separately. We process 10 seconds of audio which generates 6 seconds of synthesized audio. We apply int8 post training quantization and show both float32 and int8 latency in Table~\ref{tab:parrotron_non_stream}. In addition, we benchmark the non-streaming Griffin Lim ({\nGL} used by Parrotron model in \cite{PARROTRON_CONFORMER}) and the streaming Griffin Lim ({\sGL} used by Parrotron model in \cite{VOCODER_GL_MELGAN}). As shown in Table~\ref{tab:parrotron_non_stream}, we observe that running the model components on-device takes more than several seconds each, both when using float32 and int8 tensor operations. It makes it unusable for real STS applications(according to our user study acceptable latency has to be less around $300ms$). 

\begin{table}[t]
  \caption{10s benchmark of non-streaming Parrotron on Pixel4}
  \label{tab:parrotron_non_stream}
  \scalebox{0.9}{
      \begin{tabular}{p{2.0cm} | p{2.0cm}  | p{2.0cm} }
        \textbf{Model part} & \textbf{Latency[sec] float32} & \textbf{Latency[sec] int8} \\
        \hline
          \textit{Encoder}  & 2.8    & 2.6  \\    
          \textit{Decoder}       & 2.7    & 2.4  \\
          \textit{Vocoder sGL}    & 2.4    &   \\      
          \textit{Vocoder {\nGL}}    & 7    &   \\      
      \end{tabular}
  }
  \vspace{-2mm}
\end{table}

Non streaming \textit{Base} model has the best WER as shown on~Table~\ref{tab:parrotron_stream_wer}. It is a bit higher than WER reported in \cite{PARROTRON_CONFORMER} since we do not do extra model tuning applied in \cite{PARROTRON_CONFORMER}. We use a standard approach for all our experiments: train a model on training data, use validation data to select the best model and then test it on test data (no additional model tuning). Even though \textit{Base} model has the best WER, it has the worst \textit{total delay} of 7.9 seconds (sum of Encoder, Decoder and {\sGL} on Table~\ref{tab:parrotron_non_stream}) to process 10 seconds of input audio. For such a model \textit{total delay} goes up with the increase of the length of the input audio, but for below streaming models \textit{total delay} does not depend on input audio length.

\begin{table}[t]
  \caption{\textit{Base} model vs streaming model accuracy and \textit{total delay} of processing 10 seconds audio}
  \label{tab:parrotron_stream_wer}
  \vspace{-2mm}
  \scalebox{0.9}{
      \begin{tabular}{p{3.0cm} | p{1.2cm} | p{2.8cm} }
        \textbf{Model} & \textbf{WER[\%]} & \textbf{total delay float [ms]} \\
        \hline
          \textit{Base} \cite{PARROTRON_CONFORMER}* & 14.7     &  7900  \\
          \textit{Streaming LSA\_LS}       & 15.3    &  400 \\       
          \textit{Streaming LSA}       & 16.4    &  440    \\   
          \textit{Streaming Causal Base}       & 19.2    & 0 \\

      \end{tabular}
  }
\end{table}

We benchmarked streaming aware models presented in sections~\ref{causal}, \ref{lsa} and \ref{lsa_ls} (with streaming Griffin-Lim vocoder \cite{VOCODER_GL_MELGAN}). As expected, the streaming \textit{Causal Base} model has the best \textit{total delay} equal 0ms and the worst WER equal 19.2\%. Hybrid lookahead model \textit{LSA\_LS}(WER=15.3\%) outperforms both lookahead \textit{LSA}(WER=16.4\%) and Causal models. It proves that combination of lookahead Stacker with lookahead local self attention (\textit{LSA\_LS} model) is superior when compared to look-ahead local self attention (\textit{LSA} model) in WER metric with the same \textit{total delay} budget. Audio samples generated with these models can be found here~\cite{LINK}.

With the best streaming model \textit{LSA\_LS} the difference with non-streaming \textit{Base} model becomes 0.6\% absolute. This is while substantially reducing the total delay (down to 400ms) when compared to the \textit{Base} model 7400ms. Using this encoder with our streaming decoder and vocoder seems to build our optimal streaming-aware speech conversion system.

\subsection{Streaming decoder benchmarks} \label{decoder_experiments}

Streaming decoder {\sDec} takes data from "Encoded sequence" (as shown on Figure~\ref{fig:parrotron_inference}) then decodes it and passes it through the streaming vocoder. To stream the vocoder, we experiment with two streaming versions of vocoders: Streaming Griffin-Lim (denoted {\sGL}) and Streaming MelGAN (denoted {\sMelGAN}), described in \cite{VOCODER_GL_MELGAN}. In all our experiments, both of these vocoders look ahead into 1 hop size, hence they have a delay of 12.5ms. To optimize the latency of {\sMelGAN}, we process two frames at once, employing  streaming with external states as discussed in~\cite{KWS_STREAMING}.

In Table~\ref{tab:decoder_vocoder_stream}, we benchmark the latency, Real Time Factor (RTF is the ratio of the input duration to processing time), the TFLite model size and memory footprint when using the streaming decoder for both vocoders. To reduce the model size, we apply int8 post training quantization and report the impact on size and latency. In all these experiments, the streaming decoder generates two frames per call and the vocoder converts them into 25ms of audio.

We observe that the model size of {\sGL} is 0.1MB, which is a key advantage over {\sMelGAN} (25MB). While  {\sMelGAN} requires a larger memory footprint, it is fully convolutional, allowing us to process multiple samples per streaming inference step in parallel -- a major latency advantage over sequential vocoders, such as {\sGL} and WaveRNN~\cite{WaveRNN}. We also find that the use of int8 version of the decoder exert substantial improvements for all metrics over float32 (as shown on Table~\ref{tab:decoder_vocoder_stream}).

\begin{table}[t]
  \caption{25ms benchmark of streaming decoder {\sDec} with vocoders (\textbf{{\sGL}}, \textbf{{\sMelGAN}}) on Pixel4}
  \label{tab:decoder_vocoder_stream}
  \vspace{-2mm}
  \scalebox{0.9}{
      \begin{tabular}{p{1.8cm} | p{1.1cm}  | p{1.2cm} | p{1.1cm} | p{1.2cm} | p{1.1cm} | p{1.1cm} }
        \textbf{ } & \multicolumn{2}{|c|}{float32} & 
        \multicolumn{2}{|c|}{int8} &
        \multicolumn{2}{c}{\bfseries } \\
         & {\sDec}+\textbf{{\sGL}}    & {\sDec}+\textbf{{\sMelGAN}} & {\sDec}+\textbf{{\sGL}} &   {\sDec}+\textbf{{\sMelGAN}}   \\
        \hline
          \textit{Latency[ms]}  & 16.0    & 13.4    & 13.6    & 11.0    \\    
          \textit{RTF}          & 1.6x    & 1.9x    & 1.8x    & 2.3x    \\
          \textit{Size [MB]}    & 122     & 147     & 30      & 55    \\
          \textit{Memory[MB]}    & 139     & 166     & 44      & 64    \\
      \end{tabular}
  }
\end{table}

\subsection{Quantization of streaming model with \textit{LSA\_LS} encoder} \label{experiments}

We now explore the use of weight and activation quantization to reduce the memory footprint and latency of our best streaming model with \textit{LSA\_LS} encoder, shown on Figure~\ref{fig:streaming_encoder}. We apply int8 post training quantization~\cite{INT8_POST_TRAINING_QUANTIZATION}. To further reduce the encoder size, we perform a hybrid approach of int4 weight and int8 activation quantization aware training as in~\cite{ASR_INT4} for our encoder. We first benchmark our most accurate streaming-aware encoder \textit{LSA\_LS} on Pixel4 alone. As shown in Table~\ref{tab:encoder_stream}, the latency of processing 80ms of audio (our streaming chunk) is only 40ms using float32 and 32ms using int8 quantization.  This also leads to an encoder size of about 70MB (int4). We observe that float and quantized encoders can run in real time: RTF $\geq$ 2x.

\begin{table}[t]
  \caption{80ms benchmark of streaming encoder on Pixel4}
  \label{tab:encoder_stream}
  \scalebox{0.9}{
      \begin{tabular}{p{3cm} | p{1.2cm}  | p{2.7cm}  }
        \textbf{ } & {float32} & {int8 (int4*)} \\
        \hline
          \textit{Latency[ms]}  & 40    & 32      \\    
          \textit{RTF}          & 2x    & 2.5x    \\
          \textit{Model size [MB]}    & 436     & 111 (70* with int4)    \\
          \textit{Memory size [MB]}    & 905     & 186    \\
      \end{tabular}
  }
  \vspace{-4mm}
\end{table}

We now evaluate the impact of quantization on accuracy of our best streaming model with \textit{LSA\_LS} encoder. We report the WER for both int4 and int8 quantized \textit{LSA\_LS} encoder, while always using our int8 streaming decoder ({\sDec}),  with streaming vocoders {\sGL} and {\sMelGAN}. The results are shown in Table~\ref{tab:e2e}. Quantization helped to reduce total delay from 400ms(on Table~\ref{tab:parrotron_stream_wer}) to 320ms as shown on Table~\ref{tab:e2e}. We observe that the use of {\sGL} always exerts significantly better quality than {\sMelGAN}. We only lose 0.2\% in WER if we use int4 quantization for our encoder.  We  find that quantization of both encoder and decoder does not substantially affect quality.

\begin{table}[htbp]
  \caption{WERs of the STS conversion system with streaming quantized encoder \textit{LSA\_LS} decoder and vocoders.}
  \label{tab:e2e}
  \scalebox{0.9}{
      \begin{tabular}{p{3.0cm} | p{0.6cm}  | p{0.6cm}  | p{2.6cm}   }
        \textbf{ } & \textbf{{\sGL}} & \textbf{\sMelGAN} & {total delay int8 [ms]} \\
        \hline
          \textit{\textit{LSA\_LS} int8, {\sDec} int8}          & 15.4    & 15.9 & 320      \\
          \textit{\textit{LSA\_LS} int4, {\sDec} int8}          & 15.6    & 15.8 & 320     \\
      \end{tabular}
  }
  \vspace{-4mm}
\end{table}

\section{Conclusion}

We have presented an on-device streaming aware STS conversion model. Our proposed system consists of a streaming-aware encoder, a streaming decoder and a streaming vocoder. We have explored a novel Conformer encoder(\textit{LSA\_LS}) based on look-ahead self-attention and look-ahead stacker. It  improves quality  compared to a fully casual model and lookahead self-attention(\textit{LSA}), while also substantially minimizes the \textit{total delay} when compared to a full context non-streaming \textit{Base} model.  Studying a variety of quantization configurations, we have discussed a hybrid approach which quantizes the model weights using int4 and int8 to enable us to deploy this model locally on-device. Our best configuration runs with a Real-Time-Factor of $2{\times}$, with a delay of 320ms, to start emitting audio as soon as the speaker stops speaking, whereas the non-streaming model requires a 7-second delay.  We also obtain a significant error reduction in comparison to a fully casual model, while only  0.7\% error degradation when compared to a full context non-streaming \textit{Base} model. We suspect that this is an acceptable trade-off for deploying such a streaming STS conversion system locally on mobile device.

\bibliographystyle{IEEEtran}
{
\bibliography{mybib}
}

\end{document}